\documentclass[twoside,a4paper,14pt]{article}
\usepackage{umlaute}
\usepackage{graphicx}
\usepackage{GIFT}

\newcommand{\Gr}{Gr\"obner }
\newcommand{\Id}{\mathop{\mathrm{Id}}\nolimits}

\newcommand{\M}{\mathbf{M}}
\newcommand{\Q}{\mathbf{Q}}
\newcommand{\C}{\mathbf{C}}

\begin{document}
\thispagestyle{plain} \setlength{\parindent}{0em}
\newcounter{sep_aff}
\setcounter{sep_aff}{
4}

\title{Towards an algorithmisation of the Dirac constraint formalism\footnote{Work supported by NEST-Adventure contract 5006 (GIFT).}}
\author{
 Gerdt V\affiliation{1}
 \and Khvedelidze A\affiliation{1,}\affiliation{2,}\affiliation{3}
\and Palii Yu\affiliation{1,}\affiliation{4}
 }
\addr(1)={Laboratory of Information Technologies,\\ Joint Institute
for Nuclear Research,\\
 Dubna, 141980, Russia,\newline [gerdt, palii]@jinr.ru}
\addr(2)={Department of Theoretical Physics,\\ A.M.Razmadze
Mathematical Institute,\\
 Tbilisi, GE-0193, Georgia}
\addr(3)={School of Mathematics and Statistics, University of
Plymouth,\\
 Plymouth, PL4 8AA, United Kingdom, \\
 akhvedelidze@plymouth.ac.uk}
 \addr(4)={Institute of Applied Physics, Moldova Academy
of Sciences, \\ Chisinau, MD-2028, Republic of Moldova}

\maketitle
\begin{abstract}
Central issues of the Dirac constraint formalism are discussed in
relation to the algorithmic methods of commutative algebra based on
the \Gr basis techniques.
 For a wide class
 of finite dimensional polynomial degenerate Lagrangian
 systems, we describe an algorithmic scheme of computation of
 the complete set of constraints, their  separation into subsets of first
 and second class constraints as well as the construction of
 a generator of local symmetry transformations.
 The proposed scheme is exemplified by considering the so-called
 light-cone Yang-Mills mechanics with an $SU(2)$ gauge structure group.
\end{abstract}

\setlength{\parindent}{1em}
\clearpage
\section{Introduction}

Lagrangians used for the description of fundamental particles,
such as electrons and photons, as well as quarks and gluons have a
{\it degenerate} Hessian functions. This rather unusual property,
compared to standard Lagrangian mechanical models, profoundly
modifies the whole mathematical description of classical
evolution. It demands the physical interpretation of constrained
variables (e.g. longitudinal components of the electromagnetic
potential) and also requires the generalisation of the canonical
quantisation scheme. From the mathematical point of view the new
element of the Hamiltonian description of a degenerate Lagrangian
system is the {\em involution analysis} of the differential
equations of motion. Its pivotal ingredients in the {\em
generalized Hamiltonian dynamics}
\cite{Dirac}-\cite{HenneauxTeitelboim} are realised in the form of
the Dirac scheme to determine constraints. This is related to
~\cite{HartleyTucker,TuckerSeiler,Seiler} the {\em formal theory}
of differential equations~\cite{Pommaret}. The process of the
determining  all the {\em integrability conditions} that can not
be derived using only the algebraic operations with the existing
differential equations is just the ``reproduction'' of constraints
in the Dirac formalism. Having a complete set of constraints we
are able to identify the set of ``truly'' dynamical equations for
this involutive system and therefore finally provide a
deterministic classical evolution of the physical observables and
perform the subsequent quantization.

Effective completion to involution of systems of differential
equations needed in field theories represents a very complicated
challenge requiring sophisticated computer-algebraic
methods~\cite{CHS00}.  Similarly the generalized Hamiltonian
formalism also needs  an efficient algorithmisation and
implementation in a proper computer algebra software.

In the present paper we apply the most universal algorithmic tool
of commutative algebra, the {\it \Gr bases}~\cite{BW98}, as the
main algorithmic ingredient of the generalized Hamiltonian
dynamics for degenerate  mechanical models with polynomial
Lagrangians. In~\cite{GG99} it was already suggested to use the
\Gr bases for the computation and separation of constrains for
such models. The underlying  Dirac-\Gr algorithm is based on the
facility of the \Gr bases method to manipulate with a polynomial
in the phase variables modulo constraint manifold, and, in
particular, to check whether the polynomial vanishes on the
manifold. In the present note we propose some further algorithmic
improvements and extensions aiming  at the computational
realization of the Hamiltonian reduction of degenerate mechanical
system possessing local symmetries.

It should be noticed that constructive ideas of the involution
analysis of differential equations combined with those from the
\Gr bases technique have culminated in the concept of involutive
bases~\cite{GB98} as a special type of \Gr bases providing the
efficient involutive algorithms~\cite{G05} for construction of the
involutive as well as the  reduced \Gr bases.

The plan of this paper is as follows. We start (Section 2) with a
brief description of the main issues in the Dirac constraint
formalism that should be put into an algorithmic form suitable for
effective calculations. In Section 3 the ways to achieve this goal
for finite-dimensional mechanical systems with polynomial
Lagrangians are described. Then (Section 4) we consider the
so-called light-cone $SU(n)$ Yang-Mills mechanics as an
interesting example of constrained model for which the first
algorithmic issue of the Dirac formalism, namely, construction of
the primary constraints, can be performed for arbitrary $n$. The
remaining algorithmic issues of the Dirac formalism are
illustrated in Section 4 for this model specified~\cite{GKM02} to
the simplest nontrivial structure group $SU(2)$. Finally, in
Section 5 some conclusions are presented.

\section{The issues requiring algorithmisation}

Here we sketch briefly the basic notions and definitions from the
Dirac constraint formalism for a finite dimensional degenerate
Lagrangian system and make a list of the main procedures  requiring
an algorithmic reformulation.

Consider an $n$-dimensional mechanical system whose configuration
space is $\mathbf{R}^n $ and the Lagrangian $L(q,\dot{q})$ is
defined on a tangent space as a function  of the coordinates
$q:=q_1, q_2, \dots , q_n$ and velocities $\dot{q}:=\dot{q}_1,
\dot{q}_2, \dots , \dot{q}_n\,.$

 The Lagrangian system is called a {\em regular} one if the rank $r:={rank}\|H_{ij}\|$
 of the corresponding
 Hessian function $H_{ij}:=\partial^2 L/{\partial \dot{q}_i}{\partial \dot{q}_j}$
 is maximal
$(r=n)$.  In this case the Euler-Lagrange equations
\begin{equation}\label{Lag_eq}
\frac{\mathrm{d}}{\mathrm{d}t}\,\bigg(\frac{\partial L}{\partial
\dot{q}_i}\bigg)- \frac{\partial L}{\partial q_i}=0\,, \ \qquad \
1\leq i\leq n\,
\end{equation}
rewritten explicitly as
$$ H_{ij}\ddot{q}_j+\frac{\partial^2 L}{\partial q_j\partial \dot{q}_i}\,\dot{q}_j-
\frac{\partial L}{\partial q_i}=0\,
$$
can be resolved with respect to the accelerations ($\ddot{q}$) and
there is no {\it hidden constraints}. Otherwise, if $r<n$, the
Euler-Lagrange equations are {\em degenerate} or {\it singular}.
In this case not all differential equations (\ref{Lag_eq}) are of
 second order, namely there are $n-r$ independent equations,
Lagrangian constraints,  containing only coordinates and
velocities. Passing to the  Hamiltonian description via a Legendre
transformation
\begin{equation}
p_i:=\frac{\partial L}{\partial \dot{q}_i} \label{def_p}
\end{equation}
the degeneracy of the Hessian manifests itself in the existence of
$n-r$ relations between coordinates and momenta, the  {\em primary
constraints}
\begin{equation}\label{pr_constr}
\phi_\alpha^{(1)} (p,q)=0\,, \ \quad  \ 1\leq \alpha\leq n-r\,.
\end{equation}
Equations (\ref{pr_constr}) define the so-called {\it primary
constraints} subset $ \Sigma_1$. This definition is implicit and
therefore the first algorithmisation topic is:
\begin{quote}
{{\bf Issue I:} \it Find all primary constraints describing the
subset  $ \Sigma_1\,.$}
\end{quote}
From (\ref{pr_constr}) the dynamics is constrained by the set
$\Sigma_1\,$ and  by the Dirac prescription is governed by the
{\em total} Hamiltonian
\begin{equation}\label{def_h_t}
 H_T:=H_C+U_\alpha \phi_\alpha^{(1)}\,,
\end{equation}
which differs from the {\em canonical} Hamiltonian $H_C(p,q)=p_iq_i
- L $ by a linear combination of the primary constraints with the
Lagrange {\em multipliers} $U_\alpha$.

The next step is to analyze the dynamical requirement that
classical trajectories remain in $\Sigma_1$ during evolution,
\begin{equation}\label{cons_cond}
\dot{\phi}_\alpha^{(1)}=\{H_T,\phi_\alpha^{(1)}\}
\stackrel{\Sigma_1}{=} 0\,.
\end{equation}
In (\ref{cons_cond}) the evolutional changes are generated by the
canonical {\em Poisson brackets} with the total Hamiltonian
(\ref{def_h_t}) and the abbreviation $\stackrel{\Sigma_1}{=}$
stands for {\em a week equality}, i.e., the right-hand side of
(\ref{cons_cond}) vanishes modulo the constraints.

The consistency condition (\ref{cons_cond}), unless it is satisfied
identically, may lead either to a contradiction or to a
determination of the Lagrange multipliers $U_\alpha$ or to new
constraints. The former case indicates that the given Hamiltonian
system is inconsistent.

In the latter case when (\ref{cons_cond}) is not satisfied
identically and is independent of the multipliers $U_\alpha$ the
left-hand side of (\ref{cons_cond}) defines the new constraints.
Otherwise, if the left-hand side depends on some Lagrange
multipliers $U_\alpha$ the consistency condition determines these
multipliers, and, therefore, the  constraints set is not enlarged
by new constraints. The subsequent iteration of this consistency
check ends up with the complete set of constraints and/or
determination of some/or all Lagrange multipliers.

The number of Lagrange multipliers $U_\alpha$ which can be found
is determined by the so-called {\em Poisson bracket matrix}
\begin{equation}\label{PoissonMatrix}
\M_{\alpha \beta}:\stackrel{\Sigma}{=} \{\phi_\alpha,\phi_\beta
\}\,,
\end{equation}
where  $\Sigma$ denotes the subset of a phase space defined by the
all constraints including primary $\phi_\alpha^{(1)}$, {\it
secondary} $\phi_\alpha^{(2)}$, {\it ternary\, }
$\phi_\alpha^{(3)}$, etc., constraints, $\Phi:=(\phi_\alpha^{(1)}\,,
\phi_\alpha^{(2)}\,, \dots \,, \phi_\alpha^{(k)}\, )$
\begin{equation}\label{constr_man}
\Sigma\ :\quad \phi_\alpha (p,q)=0\,, \ \qquad\   1\leq \alpha\leq
k\,.
\end{equation}
The co-rank $s:=k-rank (\M)$ of matrix $\M$ represent the number
of  {\it first-class constraints} $\psi_1\,, \psi_2\,, \dots\,,
\psi_s\,,$ linear combinations of constraints $\phi_\alpha$
\begin{equation}\label{eq:fc}
    \psi_\alpha (p,q)=
    \sum_{\beta}\mathrm{c}_{\alpha\beta}(p,q)\,\phi_\beta\,,
\end{equation}
whose Poisson brackets are weakly zero
\begin{equation}\label{fc_cons}
\{\psi_\alpha(p,q),\psi_\beta(p,q)\} \stackrel{\Sigma}{=} 0\,\
\qquad \ 1\leq \alpha\,, \beta\leq s\,.
\end{equation}
The remaining functionally independent constraints form the subset
of {\it second-class constraints\,.}

This  method of constraints determination in the Dirac formalism
represents the particular form of {\it completion} of the initial
Hamiltonian equations to {\it involution}; the generated constraints
are nothing else than the {\em  integrability
conditions}~\cite{HartleyTucker,TuckerSeiler,Seiler}.

Therefore the second algorithmisation challenge  can be formulated
as
\begin{quote}{ {\bf Issue II:}\ \it Determine all integrability conditions and perform
their separation into first and second class conditions. }
\end{quote}

First-class constraints play a very special role in the
Hamiltonian description: they provide the basis for {\it
generators of local symmetry transformations}. The knowledge of a
local symmetry transformation is important because according to
physical requirement the physical observables  are singlets under
the gauge symmetry transformations.

So the next important algorithmisation problem is
\begin{quote}{ {\bf Issue III:}\ \it Construct the generator of local symmetry transformation
and find the basis for singlet observables. }
\end{quote}

The last problem  has direct impact on the process of {\it
Hamiltonian reduction}, that is a formulation of a new Hamiltonian
system with a reduced number of degrees of freedom but equivalent
to the initial degenerate one~\cite{Sundermeyer,GPK,Kh}. The
presence of $s$ first-class constraints and $r:=k-s$ second-class
constraints guarantees the possibility of local reformulation of
the initial $2n$ dimensional Hamiltonian system as a $2n-2s-r$
dimensional reduced Hamiltonian system (cf.~\cite{TuckerSeiler}).

Therefore, the final fourth algorithmisation  challenge  we
formulate here as
\begin{quote}{ {\bf Issue IV:}\ \it
Construct an equivalent unconstrained  Hamiltonian system on the
reduced phase space. }
\end{quote}

\section{How the algorithm works }

Here we extend the main ideas of ~\cite{GG99} and describe the
algorithmic basics that can be used to solve the problems stated
in the previous section. In doing so, we restrict our
consideration to an arbitrary  dynamical system with finitely many
degrees of freedom whose Lagrangian is a polynomial in coordinates
and velocities with rational (possibly parametric)
coefficients\footnote{Throughout this section we use some standard
notions and definitions of commutative algebra~(see, for example,
\cite{CLO}).} $L(q,\dot{q})\in \Q[q,\dot{q}]$.

\subsection{Primary constraints}

For degenerate systems the primary constraints~(\ref{pr_constr})
are consequences of the polynomial relations~(\ref{def_p}). These
relations generate the polynomial ideal in $\Q[q,\dot{q},p]$
\begin{equation}
I_{p,q,\dot{q}}\equiv\Id(\cup_{i=1}^n \{p_i-\partial L/\partial
\dot{q}_i\})\subset \Q[p,q,\dot{q}]\,. \label{Id_pqdotq}
\end{equation}
Thereby, primary constraints~(\ref{pr_constr}) belong to the
radical $\sqrt{I_{p,q}}$ of the elimination ideal
$$
I_{p,q}=I_{p,q,\dot{q}}\cap \Q[p,q]\,.
$$
Correspondingly, for an appropriate term ordering which eliminates
$\dot{q}$, a \Gr basis of $I_{p,q}$ (denotation: $GB(I_{p,q})$) is
given by~\cite{BW98,CLO}
$$
GB(I_{p,q})=GB(I_{p,q,\dot{q}})\cap \Q[p,q]\,.
$$
This means that construction of the \Gr basis for the
ideal~(\ref{Id_pqdotq}) with omitting elements in the basis
depending on velocities and then constructing of
$GB(\sqrt{I_{p,q}})$ allows us to compute the set of primary
constraints. If $GB(\sqrt{I_{p,q}})=\emptyset$ then the dynamical
system is regular. Otherwise, the algebraically independent set
$\Phi_1$ of primary constraints is the subset $\Phi_1\subset
GB(\sqrt{I_{p,q}})$ such that
\begin{equation}
\forall \phi(p,q)\in \Phi_1\ :\ \phi(p,q) \not \in
\Id(\Phi_1\setminus \{\phi(p,q) \})\,. \label{check}
\end{equation}
Verification of~(\ref{check}) is algorithmically done by computing
the following {\it normal form}: $NF(\phi,GB(\Id(\Phi_1\setminus
\{\phi \}))$.

Therefore, all the computational steps described above admit full
algorithmisation by means of \Gr bases. In addition, the canonical
Hamiltonian $H_c(p,q)$ is computed as
$NF(p_iq_i-L,GB(I_{p,q,\dot{q}}))$.

\subsection{Complete set of constraints and their separation}

The dynamical consequences~(\ref{cons_cond}) of a primary
constraint can also be algorithmically analyzed by computing the
normal form of the Poisson brackets of the primary constraint and
the total Hamiltonian modulo $GB(\sqrt{I_{p,q}})$. Here  the
Lagrange multipliers $U_\alpha$ in (\ref{def_h_t}) are treated as
time-dependent functions. If the non-vanishing normal form does
not contain $U_\alpha$, then it is nothing else than the secondary
constraint. In this case the set of primary constraints is
enlarged by  the secondary constraint obtained and the process is
iterated. At the end either the complete set $\Phi$ of
constraints~(\ref{constr_man}) is constructed or inconsistency of
the dynamical system is detected. The detection holds when the
intermediate \Gr basis, whose computation is a part of the
iterative procedure, becomes $\{1\}$.

To separate the set $\Phi=\{\phi_1,\ldots,\phi_k\}$ into of first
and second class constraints  the Poisson bracket $k\times k$
matrix $\M$~(\ref{PoissonMatrix}) is built. Its entries are
computed as normal forms of the Poisson brackets of the
constraints modulo a \Gr basis of the ideal generated by set
$\Phi$.

To construct $s:=k-r;\,$ where $r=rank(\M)$ first-class
constraints as linear combinations (\ref{eq:fc}) of constraints
(\ref{constr_man}) satisfying (\ref{fc_cons}) it suffices to find
the basis $P\:=\{p_1,\ldots,p_{k-r}\}$ of the null space (kernel)
of the linear transformation defined by $\M$. Every vector $p\in
P$ generates the first-class constraint of form $p_\alpha
\phi_\alpha$.

Now consider the $s\times k$ matrix $(p_j)_\alpha$ composed of
components of vectors in $P$ and find a basis
$T:=\{t_1,\ldots,t_{r}\}$ of the null space of the corresponding
linear transformation. For every vector $t\in T$  the second-class
constraint is constructed as $t_\alpha \phi_\alpha$.
\clearpage
Thus the constraints separation can be done using  linear algebra
operations with the matrix $\M$ alone. Together with the \Gr bases
technique this implies full algorithmisation for computing the
complete set of algebraically independent constraints and their
separation ( {\bf Issues I} and {\bf II} of Section 2).

\subsection{Generator of local  symmetry transformations}

The local symmetries are generated by first-class constraints
(cf.~\cite{HenneauxTeitelboim}) but the presence of the
second-class constraints makes realization of the symmetry
transformations very subtle. To overcome some of these
difficulties one can effectively eliminate the second class
constraints by changing  the initial Poisson bracket to the new
{\em Dirac bracket} defined as
$$
\{f,g\}_D:=\{f,g\}-\{f,\chi_\alpha\}\C^{-1}_{\alpha \beta}
\{\chi_\beta,g\}\,,
$$
where $\chi_\alpha$ $(1\leq \alpha\leq r)$ denotes the
second-class constraints, and the invertible $r\times r$ matrix
$\C_{\alpha\beta}$ is defined as
$$
\C_{\alpha \beta}: \stackrel{\Sigma}{=} \{\chi_\alpha,\chi_\beta
\}\,.
$$
Since for an arbitrary function  $f$ it follows that
$\{f,\chi_\alpha\}_D=0$  the second-class constraints can be set
to zero either before or after evaluating a Dirac bracket. This
last evaluation, modulo the constraint functions, can be performed
algorithmically exploiting the \Gr bases. After elimination of all
second-class constraints follow to the Dirac
conjecture~\cite{Dirac} the generator $G$ of local transformations
is expressed as a linear combination of all first-class
constraints
\begin{equation}\label{generator}
G= \sum_{\beta=1}^{k_1}\,\varepsilon^{(1)}_\beta
\phi^{(1)}_\beta\,+\,\sum_{\gamma=k_1+1}^{s}\,\varepsilon^{(2)}_\gamma
\phi^{(2)}_\gamma\,,
\end{equation}
and its action on phase space coordinates $(q, p)$ is given now with
the aid of the Dirac bracket
$$
\delta q_i=\{G,q_i\}_D,\qquad \delta p_i=\{G,p_i\}_D\,.
$$
In (\ref{generator}) the coefficients $\varepsilon^{(1)}_\beta\,$
and $\varepsilon^{(2)}_\gamma $ are functions of time $t$ and the
first sum includes  $k_1$ primary first-class constraints while
the second sum contains the all remaining first-class constraints.
Not all of the functions $\varepsilon$ in  (\ref{generator}) are
independent ones. Here we briefly state how following  the method
suggested in~\cite{Sossog} one  can extract the irreducible set of
functions from the set of $\varepsilon $. The total time
derivative of the gauge-symmetry generator~(\ref{generator}) is
given in terms of the Dirac bracket of $G$ and the canonical
Hamiltonian:
\begin{equation}\label{t_der_G}
\frac{dG}{dt}=\frac{\partial G}{\partial t}+\{G,H_C\}_D\,.
\end{equation}
Since the set of first-class constraints is complete,  the Dirac
bracket in the right-hand side of (\ref{t_der_G}) is
\begin{equation} \label{db_red}
\{\phi_\mu,H_C\}_D=\rho_{\mu\nu}\phi_\nu\,.
\end{equation}
The unctions $\rho_{\mu\nu}$ can be algorithmically computed by
using the \Gr bases method. To perform this computation one can
use, for example, the extended \Gr\ basis algorithm~\cite{BW93}.
Given a set of polynomials $F=\{f_1,\ldots,f_m\}\subset \Q[p,q]$
generating the polynomial ideal $\mathrm{Id}(F)$, this algorithm
yields the explicit representation
\begin{equation} \label{eq:gb}
g_i=h_{ij}\,f_j
\end{equation}
of elements in the \Gr basis $\{g_1\ldots,g_n\}$ of this ideal in
terms of the ideal generated  by polynomials in $F$. Therefore,
having computed a \Gr basis for the ideal generated by the
first-class constraints and the corresponding polynomial
coefficients for the elements in the \Gr basis as given
in~(\ref{eq:gb}), the coefficients $\rho_{\mu\nu}$  are easily
computed by reduction~\cite{BW98,CLO,BW93} of the Dirac bracket in
(\ref{db_red}) modulo the \Gr\ basis expressed in terms of the
first-class constraints $\phi_\nu$. Note that one can similarly
compute the algebra of first-class constraints
$$
\{\phi_\alpha,\phi_\beta\}_D=\varrho_{\alpha\beta\gamma}\phi_\gamma\,,
$$
if the structure functions  $\varrho_{\alpha\beta\gamma}$ are
polynomials in $p,q$.

The generator of local transformation is conserved modulo the
primary constraints
\begin{equation}\label{eq:gen}
\frac{dG}{dt} \stackrel{\Sigma_1}{=} 0\, \ \Rightarrow\
\dot{\varepsilon}^{(2)}_\gamma
\phi^{(2)}_\gamma+\varepsilon^{(1)}_\beta \rho_{\beta\gamma}
\phi^{(2)}_\gamma + \varepsilon^{(2)}_\delta \rho_{\delta\gamma}
\phi^{(2)}_\gamma \stackrel{\Sigma_1}{=} 0\,.
\end{equation}
Since, by their construction,the  constraints $\phi^{(2)}_\gamma$
do not vanish on the primary-constraint manifold $\Sigma_1$, the
relations (\ref{eq:gen}) represent the following system of
differential equations on the gauge functions
$\varepsilon^{(1)}_\beta$ and $\varepsilon^{(2)}_\gamma$
\begin{equation} \label{diff_syst}
\dot{\varepsilon}^{(2)}_\gamma +\varepsilon^{(1)}_\beta
\rho_{\beta\gamma}+\varepsilon^{(2)}_\delta
\rho_{\delta\gamma}=0\,, \quad (k_1+1\leq \gamma\leq s)\,,
\end{equation}
where the index $\beta$ runs from $1$ to $k_1$, $\gamma$ runs from
$k_1+1$ to $s$ and the functions $\rho_{\mu\nu}$ are projected on
to the subset $\Sigma_1$.

Since the differential system (\ref{diff_syst}) is
underdetermined, one can express the functions
$\varepsilon^{(1)}_\beta$ in terms of arbitrary functions
$\varepsilon^{(2)}_\gamma(t)$  and their
derivatives~\cite{Sossog}. Since this  last procedure is
algorithmic, this completes the algorithmic construction of the
generator of the local symmetry transformation.

The above described algorithmic procedures have been implemented
as a Maple package (currently for Maple 10), and this package was
used to perform the computations presented in the next section.

It is worth noting here that the remaining part of {\bf Issue III}
as well as {\bf Issue IV} still require an algorithmisation.

\section{Light-cone Yang-Mills mechanics }

Now we discuss the application of the general scheme described
above to a mechanical model originated from Yang-Mills  gauge
theory formulated on the light-cone under the assumption of
spatial homogeneity of the gauge fields.

The standard action of Yang-Mills field theory with structure
group $SU(n)$ in four-dimensional Minkowski space $M_4$, endowed
with a metric $\eta$ is
\begin{equation}
\label{eq:gaction} S : = \frac{1}{g_0^2}\, \int_{M_4} \mbox{tr} \,
F\wedge * F \,,
\end{equation}
where $g_0$ is a coupling constant and the $SU(n)$ algebra valued
curvature two-form
$$
F:= d A +  A \wedge A
$$
is constructed from the connection one-form $A$. The connection
and curvature, as Lie algebra valued quantities, are expressed in
terms of the antihermitian algebra basis $T^a $,
$$
A = A^a \, T^a \,, \qquad F = F^a \, T^a\,\,.\qquad a=1, 2, \,
\dots , n^2-1\,.
$$
The metric $\eta_{\gamma\delta}$ enters the action through the
dual field strength tensor defined in accordance with the Hodge
star operation \newline$
* F_{\mu\nu}  := \frac{1}{2}\,\sqrt{\det(\eta)}\, \epsilon_{\mu\nu\alpha\beta}\,
F^{\alpha\beta}\,, $ with totally antisymmetric tensor
$\epsilon_{\mu\nu\alpha\beta}$.

The light-cone version of the theory is formulated using the frame
where  two vectors $ e_{\pm} := \frac{1}{\sqrt 2} \, \left( e_0
\pm e_3 \right) \, $ tangent to the light-cone are combined with
the orthogonal pair $ e_k \,, k = 1, 2 \,.$ The corresponding
coordinates are usually called (see, e.g.~\cite{Brodsky})
light-cone coordinates $ x^\mu = \left( x^+, x^-, x^\bot\right) $
$$
x^\pm := \frac{1}{\sqrt 2}\, \left( x^0 \pm x^3 \right) \,, \qquad
x^\bot :=   x^k \,,\quad k = 1, 2 \,.
$$
The non-zero components of the metric are
$\eta_{+-}=\eta_{-+}=-\eta_{11}=-\eta_{22}=1\,.$ The connection
one-form in the light-cone basis is given as
\begin{equation} \label{eq:conlc}
A := A_+ \, dx^+ + A_- \, dx^- + A_k \, dx^k \,.
\end{equation}

By definition, the Lagrangian of light-cone Yang-Mills mechanics
follows from the corresponding Lagrangian of Yang-Mills theory if
one supposes that the components of the connection one-form $A$ in
(\ref{eq:conlc}) only depend on the light-cone ``time variable''
$x^+$
$$
A_\pm = A_\pm(x^+)\,, \qquad A_k = A_k(x^+) \,.
$$
Substitution of this {\em ansatz} into the classical action
(\ref{eq:gaction}) defines the Lagrangian of light-cone Yang-Mills
mechanics
\begin{equation} \label{eq:lagr}
L := \frac{1}{2g^2} \, \left( F^a_{+ -} \,  F^a_{+ -} + 2 \,
F^a_{+ k} \, F^a_{- k} - F^a_{12} \, F^a_{12} \right)\,,
\end{equation}
where $g$ is the ``renormalized''  coupling constant
$g^2=g_0^2/{(\mathrm{Volume})}$ and the light-cone components of
the field-strength tensor are given by
\begin{eqnarray*}
&& F^a_{+ -} := \frac{\partial A^a_-}{\partial x^+} + \mathrm{f}^{abc}\, A^b_+ \,  A^c_- \,,\\
&& F^a_{+ k} := \frac{\partial A^a_k}{\partial x^+} + \mathrm{f}^{abc}\, A^b_+ \,  A^c_k \,, \\
&& F^a_{- k} := \mathrm{f}^{abc} \,  A^b_- \, A^c_k \,, \\
&& F^a_{i j} := \mathrm{f}^{abc}\, A^b_i \, A^c_j\,, \quad i,j,k =
1,2 \,.
\end{eqnarray*}
Therefore,  (\ref{eq:lagr}) determines the $SU(n)$ Yang-Mills
light-cone mechanics as $4(n^2-1)$- dimensional system with
configuration coordinates $A_\pm \,,A_k$ evolving  with respect to
the light-cone time $\tau := x^+$.

The Legendre transformation
\begin{eqnarray*}
&&
\pi^+_a  :=  \frac{\partial L}{\partial \dot{A^a_+}} =0\,,\\
&& \pi^-_a  :=  \frac{\partial L}{\partial \dot{A^a_-}} =
\frac{1}{g^2} \, \left( \dot{A^a_- } + \mathrm{f}^{abc} \, A^b_+ \, A^c_- \right) \,, \\
&& \pi_a^k := \frac{\partial L}{\partial \dot{A^a_k}} =
\frac{1}{g^2} \, \mathrm{f}^{abc} \, A^b_- \, A^c_k \,
\end{eqnarray*}
gives  the canonical Hamiltonian
\begin{equation} \label{can_ham}
H_C = \frac{g^2}{2}\,  \pi^-_a  \pi^-_a - \, \mathrm{f}^{abc} \,
A^b_+ \left(A^c_- \, \pi^-_a  + A^c_k \,\pi^k_a \right) +
\frac{1}{2g^2}F^a_{12}F^a_{12}\,.
\end{equation}
The non-vanishing Poisson brackets between the fundamental
canonical variables are
\begin{eqnarray*}
\{ A^a_\pm \,, \pi^\pm_b \} = \delta^a_b \,,\qquad  \{ A^a_k \,,
\pi_b^l \} = \delta_k^l \delta^a_b \,.
\end{eqnarray*}

The Hessian of the Lagrangian system (\ref{eq:lagr}) is
degenerate, $\det ||\frac{\partial^2 L}{\partial \dot{A}\partial
\dot{A}}||= 0$, and as a result there are primary constraints
whose computation by the algorithm of Section 3.1 gives
\begin{eqnarray}
&& \varphi^{(1)}_a := \pi^+_a = 0 \,, \label{pr_c_phi} \\
&& \chi^a_k := g^2 \, \pi^a_k  + \mathrm{f}^{abc} \, A^b_- A^c_k =
0\,. \label{pr_c_chi}
\end{eqnarray}
The non-vanishing  Poisson brackets between these constraints are
\begin{eqnarray*}
\{ \chi^a_i \,, \chi^b_j\} = \, 2\,g^2\mathrm{f}^{abc} A_-^c\delta_{i
j}\,.
\end{eqnarray*}
According to the Dirac prescription, the presence of primary
constraints affects the  dynamics of the degenerate system. Now
the generic evolution is governed by the total Hamiltonian
$$
H_T := H_C + U_a(\tau)\varphi^{(1)}_a + V^a_k(\tau)\chi^a_k\,,
$$
where the Lagrange multipliers $U_a(\tau)$ and $V^a_k(\tau)$ are
unspecified functions of the light-cone time $\tau$. Using this
Hamiltonian the dynamical self-consistence of the primary
constraints may be checked. From the requirement of conservation
of the primary constraints $\varphi^{(1)}_a$ it follows that
\begin{equation} \label{phi_1}
0 = \dot \varphi^{(1)}_a = \{\pi^+_a\,, H_T\} = \mathrm{f}^{abc}
\left(A^b_- \pi^-_c  +  A^b_k \pi^k_c \right)\,.
\end{equation}
Therefore, there are three secondary constraints $\varphi^{(2)}_a$
\begin{equation} \label{phi_2}
\varphi^{(2)}_a := \mathrm{f}_{abc} \left(A^b_-  \pi^-_c  +  A^b_k
\pi^k_c \right)=0\,
\end{equation}
which obey the $SU(n)$ algebra
$$
\{ \varphi^{(2)}_a \,, \varphi^{(2)}_b \} = \mathrm{f}_{abc}\,
\varphi^{(2)}_c \,.
$$
The same procedure for the primary constraints $\chi^a_k$ gives
the following self-consistency conditions
$$
0 = {\dot\chi}^a_k = \{\chi^a_k\,, H_C \} - 2\, g^2\,
\mathrm{f}^{abc} \, V^b_k\, A^c_-   \,.
$$

A further issue, the identification of the first class constraints
among the  primary constraints $\chi^a_k$, depends on the rank of
the structure group. Below we specify to the simplest special
unitary group of rank one.

\subsection{The $SU(2)$ structure group }

Here we present the results of our computations performed for the
case of $SU(2)$ algebra where the structure constants are given by
the totally antisymmetric three dimensional Levi-Civita symbol,
$\mathrm{f}^{abc}=\epsilon^{abc}\,.$

\noindent\underline{Constraints and their separation}.
Computation of the complete set of constraints, as described in
Section 3.2, gives nine primary constraints $\varphi^{(1)}_a,
 \chi^a_{k}$  and three secondary constraints $\varphi^{(2)}_a,$ in accordance with
(\ref{pr_c_phi}) and (\ref{phi_2}). Performing the separation of
the primary constraints (\ref{pr_c_chi}) we find two additional
first-class constraints
\begin{eqnarray*}\label{psi}
\psi_k : = A^a_{-} \chi^a_k  \,,
\end{eqnarray*}
and four second class constraints
\begin{eqnarray*}\label{chi_bot}
 \chi^a_{k\bot} := \chi^a_k - \frac{\left( A^b_{-} \chi^b_k \right)
\, A^a_{-}}{(A_-^1)^2 + (A_-^2)^2 + (A_-^3)^2} \,.
\end{eqnarray*}
The new first class constraints $ \psi_i\,$ are abelian, $\{
\psi_i \,, \psi_j \} = 0 \,,$ and also have zero Poisson brackets
with all other constraints, while the second class constraints
$\chi^a_{k\bot}\,$ have the following non-zero Poisson bracket
relations
\begin{eqnarray*}
&& \{ \chi^a_{i \bot} \,, \chi^b_{j \bot} \} =
2 \, g^2 \, \epsilon^{abc} \, A^c_- \, \delta_{i j}\,, \\
&& \{\varphi^{(2)}_a \,, \chi^b_{k \bot} \} = \epsilon^{abc} \,
\chi^c_{k \bot} \,.
\end{eqnarray*}

Summarizing,  there are 8 functionally independent first-class constraints
$\varphi^{(1)}_a, \psi_k, \varphi^{(2)}_a $ and 4 second-class
constraints $\chi^a_{k \bot}$.

\noindent\underline{Generator of local symmetry
transformations}. The presence of two first class constraints
$\psi_i$ raises the question of the existence of new local
symmetries as well as  the expected $SU(2)$ gauge symmetry. To
clarify this point we construct the corresponding generator of
local symmetry transformation following Section 3.2. We start from
the expression
\begin{equation}\label{gen_su2}
G=\sum_{a=1}^{3}\,\varepsilon^{(1)}_a\varphi^{(1)}_a
+\sum_{i=1}^{2}\,\eta_{i}\psi_{i} +
\sum_{a=1}^{3}\,\varepsilon^{(2)}_a\varphi_a^{(2)}\,,
\end{equation}
with the eight light-cone time-dependent functions
$\varepsilon^{(1)}_a(\tau),\, \varepsilon^{(2)}_a(\tau)\,$ and
$\eta_{i}(\tau)\,$, then compute the functions $\rho $ (see eq.
(\ref{db_red})). Equation (\ref{eq:gen}) reads now as
$$
\left( \dot{\varepsilon}_a^{(2)}+\varepsilon_a^{(1)}
-\epsilon_{abc}\varepsilon_b^{(2)}A^{c}_{+}-\eta_{i}A^{a}_{i}
\right) {\phi}_a^{(2)}\,\stackrel{\Sigma_1}{=}\, 0\,.
$$
Therefore expressing $\varepsilon_a^{(1)}$ in terms of the
functions $\varepsilon_a^{(2)}\,,$ the generator of local
transformation takes the final form
\begin{equation} \label{gen_su2_fin}
G = \left( -\dot{\varepsilon}_a^{(2)}
+\epsilon_{abc}\varepsilon_b^{(2)}A^{c}_{+}+\eta_{i}A^{a}_{i}\right)
{\phi}_a^{(1)} +\eta_{i}\psi_{i} +
\varepsilon_a^{(2)}{\phi}_a^{(2)} \,.
\end{equation}

Analyzing the changes of the canonical coordinates $A^a$ and
$\pi^{a}$ generated by (\ref{gen_su2_fin}) we find that the
abelian subgroup of the 5-parameter local symmetry is in some
sense ``inherited'' from the rigid conformal symmetry of initial
Yang-Mills theory. But now, instead of the conformal symmetry, the
light-cone $SU(2)$ Yang-Mills mechanics has the $SL(2,R)$
dynamical group of symmetry. Moreover, the group action is
accompanied by the abelian transformations generated by two
constraints $\psi_i\,.$ A detailed discussion of this symmetry
realization will be given elsewhere.

\noindent\underline{Hamiltonian reduction to unconstrained
system}. Now that we have the generator of local transformation,
we can address the question of finding a set of suitable
coordinates part of which represent the invariants of these
transformations. Solving this problem will let us project our
system onto the constraint manifold and thus determine the
unconstrained Hamiltonian system. We refer for details to
\cite{GKM02}, and here present the set of corresponding singlet
variables (as an example of the solution of  the second part of
({\bf Issue III} ). We also give a result of subsequent
implementation of a Hamiltonian reduction ({\bf Issue IV}) of the
``redundant'' degrees of freedom associated to the symmetries
generated by constraints $\varphi^{(1)}_a$, $\varphi^{(2)}_a$ and
$\psi_a$.

Let us pass to a matrix notation: the $3\times 3$ matrix $A_{ab}$
whose entries of the first two columns are $A^a_i$ and the third
column is composed by the elements $A^a_-$. Now one can verify
that the elimination of local degrees of freedom associated with
the three constraints $\varphi^{(2)}_a$ can be achieved by using
the {\it polar representation}~\cite{Zelobenko}
$$
A = O S
$$
where $S$ is a  positive definite $3 \times 3$ symmetric matrix
and the orthogonal matrix $O$ is parameterized by three Euler
angles. It turns out that these three angles represent the pure
gauge degrees of freedom corresponding to the  constraints
$\varphi^{(2)}_a\,$.

To find the gauge degrees connected with the remaining two abelian
constraints $\psi_1\,,\psi_2$ one can pass to a principal axes
representation for the symmetric matrix $S$
$$
S = R^T\, \mbox{diag}\left( q_1\,,   q_2\,, q_3 \right)\, R\,
$$
with the orthogonal matrix $R(\chi_1,\chi_2, \chi_3)$ given in
terms of the Euler angles $(\chi_1,\chi_2,\chi_3)$. Now again it
turns out that the two angles $\chi_1$ and $\chi_2$ are pure gauge
degrees of freedom.

\noindent Solving for the remaining second class constraints
${\chi_i^a}_\bot$ leads to an unconstrained system which
represents a {\it free particle} or,  considering the complex
solutions to the second class constraints, to a more interesting
model, the so-called {\it conformal mechanics}. In this case the
diagonal variable $q_1$ and the angular variable $\chi_3$ together
with the corresponding conjugate momenta $p_1$ and $p_{\chi_3}$
are two unconstrained canonical pairs and their dynamics is
governed by the reduced Hamiltonian
\begin{equation}\label{eq:rhym}
H = \ \frac{g^2}{2}\left( p^2_1\ +
\frac{p_{\chi_3}^2}{4}\,\frac{1}{q_1^2}\right)\,,
\end{equation}
which  is a projection of the canonical Hamiltonian
(\ref{can_ham}) to the constraints shell. Finally, noting that
$p_{\chi_3}$ is a constant of motion, the Hamiltonian
(\ref{eq:rhym}) coincides with the Hamiltonian of conformal
mechanics with the coupling constant $p^2_{\chi_3}/4\,$.

\section{Concluding comments }

In this paper we have raised several issues for a constrained
mechanical systems which require  computational realization. We
described how using the \Gr basis technique  the computation and
separation of the complete set of constraints as well as the
construction of the local gauge transformations can be achieved in
degenerate mechanical models whose Lagrangians are polynomials in
coordinates and velocities. The remaining challenges, namely, the
construction  of a basis for singlet (gauge-invariant) variables
as well as the subsequent Hamiltonian reduction still needs
algorithmisation. However, a first step in this direction also has
been performed. In systems with first-class constraints the
configuration space should be factorized by the local symmetry
group in order to find a gauge invariant basis. The infinitesimal
structure of a local symmetry group is encoded in the generator of
gauge transformations, and we have shown that its construction
allows an effective algorithmisation.

As an example of the effectiveness of the proposed algorithms
light-cone Yang-Mills mechanics with the $SU(2)$ structure group
was analysed in details:  we determined and separated constraints,
constructed a local invariance transformation and found the
equivalent unconstrained Hamiltonian system.

For the $SU(2)$ light-cone mechanics the computations with our
implementation in Maple 10, which is an improved and extended
version of that given in~\cite{GG99}, takes about 1 minute on a
machine with a 1.7 GHz processor. This uses the standard \Gr
package in the Maple library. Unfortunately, recent extensions of
the Maple \Gr bases facilities with the packages {Gb} and {Fgb}
developed by J.C.~Faug\`{e}re~\cite{Faugere} do not improve on the
standard package. {Gb} is slower for our problems while Fgb cannot
deal with the parametric coefficients. For the same reason we
cannot use our software GINV~\cite{ginv} to implement the
involutive algorithms~\cite{G05} for involutive or/and \Gr bases.
Manipulation with parametric coefficients is essential for the
Dirac formalism due to the presence of physical parameters (e.g.
masses, coupling constants) in the initial Lagrangian, the
Lagrange multipliers in the total Hamiltonian (\ref{def_h_t}) and
the time-dependent functions in the generator (\ref{generator}) of
local symmetry transformations.

Consideration of light-cone mechanics with $n\geq3$ is under
current study. Here we note only that a recent paper
~\cite{GHKLMP} on geodesic motion on the $SU(3)$ group provides us
with a useful parametrization suitable for this investigation.

\section*{Acknowledgments}

The authors are indebted to  M.~Lavelle, D.~McMullan and
D.~Mla\-de\-nov for helpful discussions concerning this work.  The
presented research was partially supported by grant No.04-01-00784
from the Russian Foundation for Basic Research and grant 5362.2006.2
from the Ministry of Education and Science of the Russian
Federation.


\end{document}